\documentstyle[tighten,aps]{revtex}
\begin{document}

\baselineskip 18pt

\title{High-Pressure Low-Symmetry Phases of Cesium Halides}
\author{Marco Buongiorno Nardelli}
\address{Laboratorio INFM-TASC and BULL-Unix Competence Center,}
\address{AREA di Ricerca di Trieste, Padriciano 99, I-34012 Trieste, Italy}
\author{Stefano Baroni}
\address{Scuola Internazionale Superiore di Studi Avanzati (SISSA)
Via Beirut 2/4, I-34014 Trieste, Italy}
\author{Paolo Giannozzi}
\address{Scuola Normale Superiore (SNS)
Piazza dei Cavalieri 7, I-56126 Pisa, Italy}

\maketitle

\begin{abstract} The relative stability of different high-pressure
phases of various Cesium Halides is studied from first principles and
analyzed using the Landau theory of phase transitions. We present
results for CsI, CsBr, and CsCl up to pressures of $\approx 100\,\rm
GPa$. A cubic-to-orthorhombic transition, driven by the softening of an
acoustic phonon at the $M$ point of the Brillouin zone, is competing
with the cubic-to-tetragonal martensitic transition typical of these
compounds. The phonon softening takes place only in CsI and CsBr at a
residual volume of $V/V_0$=0.64, 0.52 respectively. A
cubic-to-tetragonal instability is found instead to occur at
$V/V_0\approx 0.54$ for all the compounds considered here. The
orthorhombic phase is stable only in CsI, whereas it is taken over by
the tetragonal one in the case of CsBr. Our analysis reveals the
essential role played by the phonon-strain coupling in stabilizing the
orthorhombic phase and in making the corresponding transition
first-order. \end{abstract}


\narrowtext

\section{Introduction}

Since the pioneering work of Madelung and Ewald,\cite{Mad+Ewa} a huge
amount of experimental and theoretical work has been devoted to alkali
halides which are the simplest and most representative ionic solids. The
interest in the high-pressure properties of these materials has been
recently renewed by the introduction of the diamond anvil cell
technology\cite{diamond-anvil} which has opened the way to exploring
previously unaccessible portions of their phase diagram. CsI is the
softest among alkali halides and it has the smallest optical gap. For
these reasons it is an ideal candidate to display a pressure-induced
band-overlap metallization which is actually found at a pressure
$P\approx 110\,\rm GPa$.\cite{Metallo} Looking for this effect, new and
unexpected crystal phases have been found to be stable in the pressure
range of a few tens GPa.

The first systematic studies of the high-pressure phases of Cesium
halides date back to 1984. From energy-dispersive X-ray diffraction
experiments Huang {\it et al.}\cite{Huang} and Knittle {\it et
al.}\cite{Knittle} were able to observe that CsI and CsBr (whose
low-pressure stable phase is the cubic B2), undergo a transition which
lowers their cubic symmetry when an applied pressure reduces their
residual volume to $v = V/V_0 \approx 0.54$, $V_0$ being the
equilibrium volume. The observed transition pressures are of $\approx
39$ and 53~GPa for CsI and CsBr respectively.\cite{Knittle} A similar
instability has been also reported in CsCl, at about 65 GPa.\cite{Huang}
This symmetry lowering manifests itself through the splitting of the
(110) X-ray diffraction line into two peaks which are interpreted as
corresponding to the (101) and (110) inequivalent reciprocal-lattice
vectors of a tetragonal structure. These authors did not observe any
detectable discontinuity in any physical observable, and they
conjectured therefore that the phase transition was of second order.
However, general group-theoretical considerations indicate that a
transition from a cubic to a tetragonal phase in which the only order
parameter is the unit cell shape, {\it i.e.} strain, cannot be of second
order.\cite{Anderson} We conclude therefore that a weak first-order
character of the transition must have escaped the analysis of the
experimental data. Claims that this tetragonal phase would undergo a
further transition to an orthorhombic structure\cite{asaumi} (space
group $D_{2h}^1$) were not confirmed by subsequent theoretical
work\cite{Born-Meyer,Christensen,Noi}.

The first-order character of the cubic-to-tetragonal transition in
Cesium halides was confirmed by the very first theoretical
investigations performed by Vohra {\it et al.}\cite{Born-Meyer} within a
Born-Meyer semiempirical approach. According to these authors, the
instability of the CsCl crystal structure is related to the softening of
the shear constant, $c_s=\frac{1}{2} (c_{11} - c_{12})$, due to the
competition between the long-range Coulomb attraction and the
short-range repulsive energy which becomes more and more important as
the pressure increases. They also found that the tetragonal structure
is stable against further orthorhombic distortions of symmetry
corresponding to the space group $D_{2h}^1$; i.e., for any given value
of $a$ and $c/a$, the minimum of the lattice energy was always found at
$b/a=1$ in the whole pressure range explored. Some investigations on CsI
have also been carried out from first principles. The work of
Christensen and Satpathy\cite{Christensen}---performed within the Linear
Muffin-Tin Orbital approximation---and that of Baroni and
Giannozzi\cite{Noi}---in which ab-initio pseudopotentials were
used---confirmed the overall picture proposed by Vohra {\it et
al}.\cite{Born-Meyer}.

In 1989, new and unexpected features emerged from the X-ray diffraction
experiments performed by Mao {\it et al.} \cite{Mao} on CsI. Using a
sophisticated experimental set-up based on a synchrotron radiation
source, they were able to identify a distortion of CsI from the cubic B2
to the hexagonal closed-packed structure passing through an intermediate
orthorhombic phase whose assigned symmetry ($C_{2v}^1$), however is
different from that guessed in Ref. \onlinecite{asaumi}. It was
observed that the (110) peak splits into three components, whereas two
additional lines appear on the low- and high-energy sides of the
triplet, corresponding to two diffraction peaks which are forbidden in
the B2 structure. This diffraction pattern can be explained by assuming
a $2\times 1$ reconstruction of the cubic cell characterized by the
gliding of one (110) plane, as depicted in Fig. 1. The onset of the
phase transition, characterized by the broadening of the diffraction
lines of the cubic structure, is observed at pressures as low as 15~GPa.
The observed splittings and peak intensities evolve continuously from 15
to 100~GPa, showing a single orthorhombic phase with variable
parameters. However, below 45~GPa the splittings are not large enough to
allow an unequivocal identification of the orthorhombic structure The
transition from the cubic to the orthorhombic phase was observed to be
continuous within the experimental resolution.

In a recent paper\cite{PRL} we have shown that the distortion leading to
the orthorhombic structure observed by Mao {\it et al.}\cite{Mao} is
driven by the softening of an $M^-_5$ phonon at the $M$ point of the
Brillouin zone (BZ). This softening is related to the ferroelastic
instability responsible for the previously assumed cubic-to-tetragonal
transition.  The shear constant is in fact proportional to the square of
the sound velocity along the $(110)$ direction for vibrations polarized
along $(1\bar 1 0)$. This observation suggests that a large reduction of
the magnitude of the shear constant could favour the softening of a
transverse phonon along the (110) direction. In fact, the gliding of the
(110) plane depicted in Fig. 1 represents a lattice distortion which is
very similar (but not strictly equal) to one of the two degenerate ionic
displacement patterns of the $M^-_5$ acoustic phonon mode. We are thus
led to identify the amplitude of this mode as the order parameter of the
transition, in Landau's sense.

In this paper we present a complete account of our previous work on the
high-pressure phases of CsI,\cite{PRL} and its extension to other Cesium
halides, CsBr and CsCl. Our approach is based on a combined use of the
Landau theory of phase transitions to classify the possible low-symmetry
phases and to determine the form of the interactions which can lead to
them, and on first-principles calculations of the relevant interaction
constants and energy differences as functions of the applied pressure.
Our calculations are based on density-functional theory to determine
crystal energies and their first derivatives with respect to atomic
displacements (forces) and macroscopic strain (stress), and on
density-functional perturbation theory to determine the second
derivatives (essentially, the phonon frequencies). In Sec. II we present
our theoretical framework and computational details. Sec. III contains
our results. Sec. IV contains the conclusions.

\section{Theory}

\subsection{Landau theory of phase transitions} In the Landau theory of
phase transitions, the relevant thermodynamical potential, $\cal F$,
describing the relative stability of two phases is expressed as a power
series of the so called order parameter, $u$, whose value is different
from zero in the low-symmetry (high-pressure, in the present case)
phase, while $u$ vanishes in the high-symmetry (low-pressure) phase:
\begin{equation} {\cal F} (u) = {\cal F}_0 + A_2 u^2 + A_3 u^3 + A_4 u^4
+ \cdots, \label{eq:landau} \end{equation} where the linear term in $u$
is missing to ensure the stability of the low-pressure phase. Let us
suppose for the moment that the fourth-order coefficient, $A_4$, is
positive so that it can ensure the global stability of the system, and
let us neglect all higher-order terms. If
$A_2 > 0$, the relative stability of the low- and high-symmetry phases
is determined by the sign of the discriminant $\Delta_6 = A_3^{~2}-4
A_2A_4$. If $\Delta_6 <0$ the high-symmetry ($u=0$) phase is stable,
while for $\Delta_6>0$ Eq. (1) has a minimum for $u\ne 0$ and the
low-symmetry phase prevails. In this case, the transition from the high-
to the low-symmetry phase is discontinuous, and it is said to be {\em
first-order}. Continuous {\em second-order} transitions are possible if,
because of symmetry, the third-order coefficient vanishes identically:
$A_3 \equiv 0$. In this case the high- or low-symmetry phases are stable
according to whether the second-order coefficient, $A_2$, is positive or
negative. Suppose now that $A_4$ is negative in the region of the phase
diagram where $A_2$ softens and that $A_5\equiv 0$. In this case, the
global stability of the system must be ensured by the sixth- and
higher-order terms and a first-order transition to the low-symmetry
phase would then occur when $\Delta_8 \equiv A_4^{~2}-4A_2A_6=0$.

For structural transitions where the order parameter is the amplitude of
a lattice distortion, $A_2$ is essentially the square of the phonon
frequency associated to that distortion. In the present case, the phonon
which goes soft is the $M^-_5$ mode at the $M$ point of the BZ. The
small group of the $M$ point---whose coordinates are $({1 \over 2}{1
\over 2} 0)$---is $D_{4h}$ and its star is made of three equivalent
points.  $M_5^-$ transforms according to a doubly degenerate irreducible
representation of $D_{4h}$, so that the order parameter associated with
this mode is six-dimensional: ${\bf u}\equiv \{ {u}_i \}_{i=1,6}$. The
$M^-_5$ representation is odd with respect to inversion, so that no
third-order (or, more generally, odd-order) terms must be present in Eq. (1)
in order to ensure invariance with respect to the (cubic) symmetry group
of the low-pressure phase. Following Landau's argument, we conclude that
$A_3\equiv 0$, and $A_5\equiv 0$. Assuming that the sixth- and
higher-order terms in Eq. (1) are positive in the neighborhood of the
transition, this is then first- or second-order, according to whether
the fourth-order coefficient is negative or positive, respectively.

\subsection{First-principles techniques} In this paper we will ignore
any entropy effects and restrict ourselves to zero temperature. In this
condition, the implementation of the above phenomenological scheme
requires the knowledge of the system energy as a function of the atomic
displacements (phonon amplitudes) and unit-cell distortions (strain).
Density-Functional Theory (DFT) is a very reliable and computationally
viable tool for the study of the energetics of simple materials. We
refer the reader to the many available review papers on this
subject\cite{DFT-reviews} and we summarize here the main points. The
local-density approximation (LDA) to DFT provides a practical, though
approximate, way to map the ground-state properties of a system of
interacting electrons onto those of a system of noninteracting
electrons, subject to an effective external field which depends
self-consistently on the ground-state electronic density.  DFT-LDA
allows to calculate in a rather accurate and efficient way the energies
of materials within the Born-Oppenheimer approximation, along with their
first derivatives with respect to external parameters, such as e.g.
nuclear coordinates (i.e. atomic forces)\cite{HF} and unit-cell shape
and volume (i.e. stress).\cite{Nielsen} By linearizing the DFT-LDA
equations with respect to the strenght of an external perturbation, the
second derivatives of the energy can be conveniently obtained by solving
a suitable set of linear equations.\cite{DFPT,DFPT-sega} This is the essence of
the Density-Functional Perturbation Theory (DFPT) approach to lattice
dynamics, which we will employ to determine the phonon spectra of the
system considered, as functions of the crystal volume, i.e. of the
applied pressure.

Our calculations are performed in the framework of the plane-wave
pseudopotential method.\cite{DFT-reviews} For Cs and I we have used the same
pseudopotentials as in Ref. \onlinecite{Noi}.
Pseudopotentials for Br and Cl have been generated
using the method originally proposed by von
Barth and Car.\cite{pap_pasquale} For Cs, electrons up to $4d$ have been
treated as frozen in the core. The inclusion of the Cs $5s$ and $5p$
states into the valence shell is necessary to obtain sensible results.
In fact we found that a Cs pseudopotential which sustains only the $6s$
electron as a valence electron yields a monotonic decrease of the
crystal energy at decreasing volume, with no stability at all. An
explicit account for the non-linear core-corrections,\cite{Louie-NLCC}
suggested by the large size of the Cs core, does not substantially
improve the results. The inclusion of Cs $5s$ and $5p$ orbitals in the
valence does not increase dramatically the numerical burden. In fact,
the spatial extension of these orbitals is comparable to those of the
anion $s$ and $p$ valence wavefunctions, so that the plane-wave basis
set necessary to describe the latter is also adequate for the former.

Plane waves up to a kinetic energy cutoff of 25 Ry were included in the
basis set. The electron-gas exchange-correlation energy and potential
used to implement LDA are those determined by Ceperley and Alder
\cite{Ceperley} as interpolated by Perdew and Zunger.\cite{Perdew}
Brillouin-zone integrations have been performed using sets of special
points corresponding to the (666) Monkhorst-Pack mesh.\cite{MP}
This mesh---which corresponds to 10, 12, and 24
points in the cubic, tetragonal, and orthorombic structures
respectively---has been explicitly checked for convergence for all the
relevant structural properties of interest here.

\section{Results}

\subsection{Zero-pressure properties} The low-pressure structural
properties of the materials investigated have been determined by fitting
the calculated crystal energies to the Murnaghan equation of
state.\cite{Murnaghan} DFPT allows a straightforward determination of
the electronic contribution to the static dielectric constants and Born
effective charges.\cite{DFPT} The theoretical predictions for these data
are compared with experiments in Table \ref{t:parret}. The resulting
accuracy on the predicted equilibrium lattice constants and bulk moduli
is of the order of $3\%$ and 10\% respectively, i.e. in the typical
range of DFT-LDA calculations. Dielectric constants are systematically
overestimated by $\approx 20\%$, a well known drawback of the
LDA.\cite{eps-LDA} Using the theoretically determined equilibrium
lattice parameters, the phonon dispersions of the three salts have been
calculated by DFPT along the main symmetry directions. In Fig.
\ref{f:fig2} we display the calculated dispersion of CsI and compare
them with neutron-diffraction data. Selected frequencies for CsI, CsBr,
and CsCl are reported and compared with experiments in Table
\ref{t:phonon}. The agreement is quite satisfactory and gives us
confidence in the predictive power of our calculations.

\subsection{Phonons at high pressure} In order to substantiate the
hypothesis of a mode softening occurring upon increasing pressure along
the (110) acoustic branch, we have calculated the corresponding phonon
dispersion for several values of the crystal residual volume, $v$.
In Fig. \ref{f:fig3} we display our results for CsI. The $M_5^-$
acoustic phonon softens at $v^* = 0.638$, corresponding to a
pressure $P^* \approx 23~ \rm GPa$, as a consequence of the incipient
softening of the sound velocity of one of the transverse branches. The
same behavior is observed in CsBr at $v^* = 0.520$ while CsCl does not
display any tendency to softening upon increasing pressure.  The
dependence of the $M_5^-$ acoustic phonon frequency upon the residual
volume is displayed in Fig. \ref{f:fig4} for CsI, CsBr and CsCl.

\subsection{High-pressure phases} \subsubsection{Tetragonal phase} As a
first step we have reinvestigated the cubic-to-tetragonal transition
described in the introduction. In
Fig. \ref{f:tet1} we display the zero-temperature enthalpies {\it vs.}
the $c/a$ cell parameter calculated at constant pressure for CsI, CsBr
and CsCl. In CsI and in CsBr the existence of a first-order transition
is clearly visible with a second minimum occurring at $c/a > 1$ above
some critical pressure, $P_{tet}$, and separated from the high-simmetry phase
by a well defined enthalpy barrier. The transition pressures are $P_{tet} =
44$ and $58$ GPa for CsI and CsBr respectively, with a very small volume
discontinuity ($\approx 0.5\%$ and $\approx 0.7\%$ respectively). In
Fig. \ref{f:tet2} we report the $c/a$ parameter {\it vs.} $P/P_{tet}$ for
CsI and CsBr. In CsCl the stability of the low-simmetry phase cannot be
unambiguously established within our present accuracy. An instability
towards the tetragonal phase is signalled by the flatness of the
enthalpy as a function of $c/a$ for pressures $\approx 70~\rm GPa$. At
higher pressures, however, this instability seems to weaken, and
calculations for pressures as high as 100 GPa do not give any evidence
of a stable tetragonal phase.

\subsubsection{Orthorhombic phase} As we have seen in Sec. IIA, the
vanishing of the $M^-_5$ phonon frequency signals the onset of a phase
transition whose order parameter is the phonon amplitude and which can
be second-order because no third-order terms are present in the
expansion of the crystal energy in powers of the order parameter. In
order to investigate the character of the transition and determine the
possible low-symmetry phases, it is necessary to find out the expression
of the crystal energy in powers of $u$ to the lowest
meaningful order, i.e. $n=4$. In Table \ref{t:phonq} we display the
wavevectors and the polarizations corresponding to one possible choice
for the components of the six-dimensional order parameter. The
corresponding fourth-order invariant polynomials can be obtained by
using standard group theoretical techniques. The $u$'s realize an irreducible
representation of the symmetry group of the high-symmetry phase (space
group $O_h^1$).  The number of fourth-order invariants is equal to the
number of times the identical representation of $O_h$ is contained in
the symmetric part of $[ M^-_5]^4$.  A simple exercise shows that this
number is equal to four. A straightforward way to obtain them is to
symmetrize all the possible translationally invariant fourth-order
monomials with respect to $O_h$.  In order to satisfy translational
invariance, the starting monomials, $u_iu_ju_ku_l$ must satisfy the
relation: $ {\bf q}_i + {\bf q}_j + {\bf q}_k + {\bf q}_l = {\bf G}$,
where ${\bf q}_i$ is the wavevector associated to the $i$-th component of the
order parameter, and {\bf G} is a reciprocal-lattice vector of the cubic
phase. The four independent invariants are summarized in Table
\ref{t:inv}. The resulting expansion of the crystal energy
reads: \begin{equation} {\cal F}({\bf u}) = {\cal F}_0 + A_2
\sum_{i=1,6} u_i^{~2} + \sum_{i=1,4}A_4^i\,P_4^i({\bf u}) + {\cal
O}(u^6). \label{eq:free_en} \end{equation}

The possible low-symmetry stable phases compatible with this expression
for the crystal energy have been worked out in Ref. \onlinecite{Kim}.
Following the notation of Ref. \onlinecite{Kim}, these stable phases can
be classified as in Table \ref{t:P}, where each configuration
corresponds to a particular direction in the six-dimensional order
parameter space, {\it i.e.} to a particular combination of phonon
displacements. In the specific case of CsI, the four fourth-order
coefficients appearing in Eq. (2) have been fitted to the results of
DFT-LDA calculations performed for lattice distortions along some of the
order-parameter directions listed in Table \ref{t:P}. In fact, the
coefficients of the crystal energy expansion are easily obtained through
total energy and stress calculations.  With the numerical values so
obtained, we have verified that at the crystal volume where the $M^-_5$
phonon frequency softens, the only minimum of the energy given by Eq.
(2) occurs along the $P11$ direction, whereas directional minima along
all the other directions are actually saddle points. This observation
seems to suggest that a second-order transition to a phase of
tetrahedral symmetry ($T^5$) would occur at the softening pressure of
the $M^-_5$ phonon. This finding is at variance with experimental data
which indicate that the high-pressure phase has an orthorhombic
symmetry, compatible with a distortion along the $P1$ line.

The above considerations hold in the hypothesis that the strain state of
the crystal is constant across the transition, except for the isotropic
compression due to the application of a hydrostatic pressure. We have
seen, however, that the softening of the $M^-_5$ phonon mode is closely
related to the softening of the shear constant of the crystal, so that a
strong coupling between the soft mode and macroscopic strain (i.e.,
between zone-center and zone-border acoustic phonons) is to be expected.
Let us consider now the expression of the Landau crystal energy up to
fourth order in the phonon amplitude, including the coupling with the
anisotropic components of the strain tensor, $\epsilon$:
\begin{equation} {\cal F}({\bf u},\epsilon) = {\cal F}({\bf
u},\epsilon=0) + Q(u,\epsilon), \label{eq:strain} \end{equation} where
$Q(u,\epsilon)$ is a polynomial in $u$ and the components of the strain
tensor containing all the possible products up to the fourth order:
$\epsilon^2$, $u^2\epsilon$, $u^2\epsilon^2$, $\epsilon^3$, and
$\epsilon^4$. All odd terms in $u$ must vanish by symmetry. Up to fourth
order only $u^2\epsilon$ must be considered; in fact the actual value of
$\epsilon$ is the one which minimizes Eq. (\ref{eq:strain}) for a given
value of $u$, i.e. $\epsilon \propto u^2$ so that $u^2\epsilon^2$ and
$\epsilon^3$ are actually of sixth order in $u$ while $\epsilon^4$ is of
eighth order. To find the form of the $u^2\epsilon$ invariants we have
first determined all the second order monomials, $u_iu_j$, such that
${\bf k}_i + {\bf k}_j = {\bf G}$. There are nine such monomials which
realize a reducible representation of the $O_h$ point group. The
decomposition of this reducible representation into irreducible
components is displayed in Table \ref{t:irrep}. The six independent
components of the strain tensor split into three irreducible
representation of the cubic group, as indicated in Table
\ref{t:irrep_strain}. The desired invariants are finally obtained by
coupling each irreducible representation from the $uu$ set (Table
\ref{t:irrep}), with an equal irreducible representation from the
$\epsilon$ set (Table \ref{t:irrep_strain}), and listed in Table
\ref{t:inv_strain}. The resulting expression for the crystal energy
including the coupling with the strain reads: \begin{eqnarray} \quad
{\cal F}&(&{\bf u},\epsilon) = {\cal F}_0 + A_2 \sum_{i=1,6}u_i^{~2}
+\sum_{i,j=1}^6 c_{ij}\epsilon_i\epsilon_j \nonumber \\ && \quad+
\sum_{i=1,4}A_4^i\,P_4^i({\bf u}) + \sum_{i=1,4} B^i Q^i_4({\bf
u},\epsilon) + {\cal O}(u^6), \label{eq:phon_strain} \end{eqnarray}
where $c_{ij}$ are the elastic constants, $\epsilon_i$ components of the
strain tensor, and the $Q_4^i$ are the polynomials listed in Table
\ref{t:inv_strain}.

Numerical values of the $B$ coefficients have been determined along
lines similar to those used to determine the $A_4$'s. The $P11$ minimum
previously found to be stable ignoring the phonon-coupling strain is
unaffected by such a coupling, while the minima along four out of the
remaining six directions of Table \ref{t:P} are slightly modified, still
maintaining the saddle-point character they had in absence of such a
coupling. The coupling to the strain changes instead the character of
the directional minima along $P1$ and $P2$. Let us consider the simpler
case of the P1 minimum. Projecting Eq. (\ref{eq:phon_strain}) along $P1$
one obtains: \begin{eqnarray}{\cal F}^{P1}(u,\epsilon_s,\epsilon_4) &=&
A_2 u^2 + A^{P1}_4 u^4 + {1\over 2} c_s \epsilon_s^2 + {1\over 2} c_{44}
\epsilon_4^2 + \nonumber \\ &&\quad ( B^{P1}_s \epsilon_s + B^{P1}_4
\epsilon_4 ) u^2 + {\cal O}(u^6), \label{H_u_eps} \end{eqnarray} where
$u=u_1$ is the amplitude of the phonon displacement along $(1\bar 10)$,
$B^{P1}_s ={1\over 4}B^1 -2B^2$, $B^{P1}_4 = B^4$,
$\epsilon_s=\frac{1}{2}(\epsilon_{xx}+ \epsilon_{yy} -2 \epsilon_{zz})$,
and $\epsilon_4=\epsilon_{xy}$. By eliminating $\epsilon_1$ and
$\epsilon_2$ from Eq. (\ref{H_u_eps}) by the equilibrium condition
$\partial {\cal F}^{P1}/\partial\epsilon_s = \partial {\cal
F}^{P1}/\partial\epsilon_4 = 0$, one obtains: \begin{equation}
\epsilon_s = -\frac{B^{P1}_s u^2}{c_s};\quad \epsilon_4 =
-\frac{B^{P1}_4 u^2}{c_{44}}, \label{e:eps_u} \end{equation}
\begin{eqnarray} \quad \widetilde {\cal F}^{P1}(u) &=& A_2 u^2 +
\nonumber\\ \quad &+& \left ( A_4^{P1} -\frac{{B^{P1}_s}^2}{2c_s} -
\frac{{B^{P1}_4}^2}{2c_{44}} \right )u^4 + {\cal O} (u^6).
\label{H_of_u} \end{eqnarray} Eq. (\ref{H_of_u}) shows that the coupling
between the soft phonon and macroscopic strain renormalizes the
fourth-order coefficient, making it large and {\it negative}, whenever
$c_s$ or $c_{44}$ are small enough. Due to the ongoing softening of the
shear constant, $c_s$, we find that at the softening pressure the
fourth-order coefficient is negative, and we conclude that the
transition must then be first-order, occurring at a somewhat lower
pressure (see the discussion at the end of Sec. IIA). A similar
behaviour is observed along P2. A thorough study of the transition might
be performed by considering the expansion of the crystal energy up to
sixth order in the order parameter, and by fitting the relevant
coefficients to first-principles calculations, as it was done when the
coupling to the strain was neglected. However the presence of many
sixth-order invariants and the smallness of the computed energy
differences make such a procedure unpractical. We have preferred instead
to concentrate on the $P1$ line which is geometrically simpler and whose
symmetry is compatible with the observed x-ray diffraction pattern of
the high-pressure phase. For this direction we have performed straight
energy minimizations with respect to $u$ and $\epsilon$. By repeating
such a minimization for different volumes, one directly obtains the
equation of state of the crystal in the low-symmetry phase.

\subsubsection{Relative stability} CsI and CsBr both display a softening
of the $M^-_5$ phonon frequency and the ferroelastic instability
responsible for the cubic-to-tetragonal transition, whereas CsCl does
not show any tendency to phonon softening and a very weak tendency
towards the ferroelastic instability. We conclude that the
high-pressure stable phase of CsCl in the pressure range up to $\approx
100\,\rm GPa$ is cubic B2, whereas the relative stability of the
tetragonal and orthorhombic phases must be checked for CsBr.

We have performed complete energy minimizations with respect to atomic
displacements and strain parameters for both the tetragonal and the
newly determined orthorhombic structure. The structural parameters of
the orthorombic phase of CsI as functions of pressure/volume are shown
in Fig. \ref{f:fig5}. All these, including volume, are discontinuous at
the transition pressure as a consequence of its first-order character.
The volume discontinuity is however very weak ($\approx 0.1\%$) and it
is not visible on the scale of the figure. Note also the different
magnitude of Cesium and Iodine displacements and the weak
dependence of $b/a$ upon applied pressure. The different behavior of
$c/a-1$ and $b/a-1$ with respect to pressure is due to the fact that the
former is inversely proportional to the nearly vanishing shear constant,
$c_s$, whereas the latter depends on $c_{44}$ which is regular in this
pressure range. We remark that, due to the different amplitude of Cs and
I displacements in the unit cell, the space group of our final structure
is $D_{2h}^5$, different from the $C_{2v}^1$ originally proposed by Mao
{\it et al.} \cite{Mao}. However, the diffraction pattern corresponding
to the two groups is the same, and our proposed structure is
compatible with the experimental data.

We have compared the enthalpies of the various phases and identified the
transition pressures through the Maxwell construction. Results of this
procedure are shown in Fig. \ref{f:fig6} where the enthalpies of CsI and
CsBr are shown for the tetragonal and orthorhombic phases relative to
the cubic one. In agreement with experimental findings the high-pressure
stable phase of CsI is the orthorhombic one (for pressures larger than
$\approx 21\,\rm GPa$). At variance with the case of CsI, in CsBr the
tetragonal structure turns out to be more stable all over the explored
pressure range. The phonon softening is preceeded by the
cubic-to-tetragonal transformation at $P_{tet} = 58\,\rm GPa$, and we have
verified that the tetragonal phase is always favored.
So we can definitely
assign the high-pressure phase of CsBr to be tetragonal.
The cubic-to-tetragonal transition has been
experimentally observed in CsBr at $53\pm 2\,\rm GPa$ corresponding to
a residual
volume $v_{tet} \approx 0.54$.\cite{Huang}
As we have
seen, CsCl does not manifest any phonon softening upon increasing
pressure. This compound shows only a weak tendency to
a tetragonal instability at $P_{tet} \approx
69\,\rm GPa$, but the tetragonal phase seems not to be stable at higher
pressures.
Our results are in agreement with this
measurement. In CsCl, a weak tetragonal instability has been observed
experimentally at $65 \pm 5$ GPa, but an extensive study at higher
pressure is still lacking.

\section{Conclusions}

In this paper we have shown that a combined use of the Landau theory of
phase transitions and accurate first-principles calculations based of
DFT-LDA provide a reliable scheme for predicting the structural
properties of ionic systems at high pressures. In the specific case of
Cesium halides, we have observed a trend in the high-pressure phases
which can be related with the polarizability of the anion. Our results
indicate that the more polarizable is the anion the larger is the
tendency towards a tetragonal instability of the cubic B2 structure
(passing from CsCl to CsBr) and, for even larger polarizability towards
an orthorombic structure (from CsBr to CsI). Some investigations which
we have performed also for other alkali halides indicate that when the
cation is lighter (and hence less polarizable) than Cs, the stable
high-pressure phase seems always to be the cubic B2 one.

\section{Acknowledgements}

We would like to thank S. de Gironcoli and P. Pavone for valuable
discussions. Most of this work was completed while MBN was at SISSA.

\newpage

\narrowtext
\begin{table}
\caption{Comparison between theoretical and experimental lattice
constant, $a_0$, bulk modulus, $B_0$, and electronic static dielectric
constant, $\epsilon_\infty$.}
\begin{tabular}{lllll}
           &       &  $a_0$~[\AA] &  $B_0$~[GPa]  &  $\epsilon_\infty$
\\ \tableline
CsI        &{\it theory}&  4.44~~~~  &  15.4~~~~~  & 3.68 \\
           &{\it expt.} & 4.56\tablenotemark[1]
{}~~~& 13.5\tablenotemark[1] & 3.02\tablenotemark[5]    \\
\\
CsBr       &{\it theory}&  4.17~~~~     &  20.7~~~~~& 3.36   \\
           &{\it expt.} & 4.28\tablenotemark[2]~~~  & 17.9\tablenotemark[2]~~~~
 & 2.83\tablenotemark[5]  \\
\\
CsCl       &{\it theory}&  3.99~~~~     &  24.3~~~~~ & 3.16 \\
           &{\it expt.} & 4.12\tablenotemark[3]~~~  & 22.9\tablenotemark[4]~~~
& 2.67\tablenotemark[5]
\end{tabular}
\tablenotetext[1]{Ref. \protect\onlinecite{Mao}}
\tablenotetext[2]{Ref. \protect\onlinecite{Reinitz}}
\tablenotetext[3]{Ref. \protect\onlinecite{Wyckoff}}
\tablenotetext[4]{Ref. \protect\onlinecite{Singh}}
\tablenotetext[5]{Ref. \protect\onlinecite{LowMar}}
\label{t:parret}
\end{table}

\widetext

\begin{table}
\caption{Comparison between calculated and experimentally observed phonon
frequencies of Cesium halides at zero pressure, at the $\Gamma$, $M$, $X$, and
$R$
points of the BZ. Units are THz. The uncertainty on the last
figure in the experimental data is indicated in parenthesis.
}
\label{t:phonon}
\begin{tabular}{llllllllllllll}
&         & $\omega_{TO}(\Gamma)$ &$\omega_{LO}(\Gamma)$ & $\omega_{TA}(M)$
& $\omega_{LA}(M)$ & $\omega_{TO}(M)$ & $\omega_{LO}(M)$ & $\omega_{TA}(X)$
& $\omega_{LA}(X)$ & $\omega_{TO}(X)$ & $\omega_{LO}(X)$ & $\omega_{LA}(R)$
& $\omega_{LO}(R)$ \\ \tableline
CsI  & {\it theory} & 2.09 & 2.85 & 1.17 & 1.30 & 1.48 &
2.35 &1.38 &1.59 &2.18 &2.21 &1.71 &2.00 \\
     & {\it expt. } & 1.91(5)\tablenotemark[1] & 2.74(1)\tablenotemark[2]
& 1.29(6)\tablenotemark[1] &
1.34(5)\tablenotemark[1] &  ~~--         & 2.26(5)\tablenotemark[1]
&1.21(5)\tablenotemark[1] &1.30(6)\tablenotemark[1] &2.07(5)\tablenotemark[1]
 &2.16(6)\tablenotemark[1] &1.75(5)\tablenotemark[1] &~~--  \\
CsBr & {\it theory} & 2.44 & 3.53 & 1.27 & 1.48 & 1.98 &
2.79 &1.47 & 2.15 & 2.42 & 3.00 & 1.89 & 2.63 \\
     & {\it expt. } & 2.29(2)\tablenotemark[3] &  ~~--
& 1.22(3)\tablenotemark[3] &
1.61(3)\tablenotemark[3] & 1.77(5)\tablenotemark[3]
& 2.69(5)\tablenotemark[3] & 1.32\tablenotemark[3]
& 1.91\tablenotemark[3] & ~~-- & ~~-- & 1.89\tablenotemark[3]
& 2.40\tablenotemark[3] \\
CsCl & {\it theory} & 3.33 & 5.06 & 1.24 & 1.86 & 3.18 &
4.04 &1.43 &2.44 &3.21 &4.50 &1.91 &3.94  \\
     & {\it expt. } & 3.17(2)\tablenotemark[4]
&  ~~--  & 1.19(3)\tablenotemark[4] &
1.90(2) & 2.77(2)\tablenotemark[4] & 3.75(3)\tablenotemark[4]
&1.31\tablenotemark[4] &2.98\tablenotemark[4] &~~-- &~~--
&2.02\tablenotemark[4] &3.85\tablenotemark[4]
\end{tabular}
\tablenotetext[1]{Ref. \protect\onlinecite{Buhrer}}
\tablenotetext[2]{Ref. \protect\onlinecite{Lowndes}}
\tablenotetext[3]{Ref. \protect\onlinecite{Rolandson}}
\tablenotetext[4]{Ref. \protect\onlinecite{ASWW}}
\end{table}

\narrowtext

\begin{table}
\caption{Phonon wavevectors, $\bf q$, and polarizations,
$\bf e$,
of the six components of the order parameter, $\bf u$.}
\begin{tabular}{lccccccc}
 && $u_1$ & $u_2 $& $u_3$ & $u_4$ & $u_5$ & $u_6$ \\
\tableline
$\bf q$ &&
$(0{1\over 2}{1\over 2})$ &
$(0{1\over 2}{1\over 2})$ &
$({1\over 2}0{1\over 2})$ &
$({1\over 2}0{1\over 2})$ &
$({1\over 2}{1\over 2}0)$ &
$({1\over 2}{1\over 2}0)$ \\
\noalign{\vskip 3pt}
$\bf e$ &&
$(100)$ &
$\left(0{1\over\sqrt{2}} \bar{1\over\sqrt{2}} \right)$ &
$(010)$ &
$\left({1\over\sqrt{2}}  0 \bar{1\over\sqrt{2}} \right)$ &
$(001)$ &
$\left({1\over\sqrt{2}} \bar{1\over\sqrt{2}} 0 \right)$
\end{tabular}
\label{t:phonq}
\end{table}

\begin{table}
\caption{Fourth-order invariant polynomials which
enter the free energy expansion, Eq. \protect\ref{eq:free_en}.}
\begin{tabular}{cc}

$P_4^1({\bf u})$
  & $ u^4_1 + u^4_2 + u^4_3 + u^4_4 + u^4_5 + u^4_6  $ \\
$P_4^2({\bf u})$
  & $ u^2_1u^2_2 + u^2_3u^2_4 + u^2_5u^2_6 $  \\
$P_4^3({\bf u})$
  & $ -u_1u_2(u^2_3 + u^2_4 - u^2_5 - u^2_6 ) +$ \\
  & $u_3u_4(u^2_5 + u^2_6 - u^2_1 - u^2_2 ) +$   \\
  & $ u_5u_6(u^2_1 + u^2_2 - u^2_3 - u^2_4 ) $   \\
$P_4^4({\bf u})$
& $ -u_1u_2u_3u_4-u_1u_2u_5u_6
      +u_3u_4u_5u_6 $ \\
\end{tabular}
\label{t:inv}
\end{table}

\begin{table}
\caption{Classification of the possible stable phases along particular
directions of the order-parameter space.}
\begin{tabular}{lcc}
   & Direction  & Space group      \\ \tableline

P1 & (~1,~0,0,~0,~0,~0) & $D_{2h}^5$    \\
P2 & (~1,-1,0,~0,~0,~0) & $D_{2h}^{19}$ \\
P6 & (~1,~0,1,~0,~1,~0) & $C_{3v}^5$    \\
P7 & (~0,-1,0,~1,~0,~1) & $D_3^7$       \\
P9 & (~1,-1,1,-1,~0,~0) & $D_{4h}^{17}$ \\
P10& (~1,-1,0,~0,~1,-1) & $D_{4h}^{17}$ \\
P11& (~1,-1,1,~1,~1,~1) & $T^5$         \\
\end{tabular}
\label{t:P}
\end{table}

\begin{table}
\caption{Decomposition of the set
of lattice-periodic quadratic monomials in
the order parameter into irreducible representations of the cubic group.}
\begin{tabular}{cc}
Label      & Basis functions \\ \tableline

$\Gamma^+_2$ & $-u_1u_2 +u_3u_4 +u_5u_6$  \\
$\Gamma^+_3$ & $-2u_1u_2 -u_3u_4 -u_5u_6;$ \\
& $u_3u_4 -u_5u_6$             \\
$\Gamma^+_1$ & $u^2_1 +u^2_2 +u^2_3 +u^2_4 +u^2_5 +u^2_6$     \\
$\Gamma^+_3$ & $2 (u^2_1 +u^2_2) -u^2_3 -u^2_4 -u^2_5 -u^2_6;$ \\
& $u^2_3 +u^2_4 -u^2_5 -u^2_6$                   \\
$\Gamma^+_5$ & $u^2_1 -u^2_2;~ u^2_3 -u^2_4;~ u^2_5 -u^2_6$     \\
\end{tabular}
\label{t:irrep}
\end{table}

\begin{table}
\caption{Decomposition of the set of independent components of a
symmetric rank-2 tensor into irreducible representations of the cubic
group.}
\begin{tabular}{cc}
Label      & Basis functions \\ \tableline
$\Gamma_1^+$ & ${1\over 3}(\epsilon_{xx}+\epsilon_{yy}+\epsilon_{zz})$ \\
$\Gamma_3^+$ & $\epsilon_{xx}-\epsilon_{yy}$;
               $2\epsilon_{zz}-\epsilon_{xx}-\epsilon_{xx})$ \\
$\Gamma_5^+$ & $\epsilon_{xy};\,\epsilon_{yz};\,\epsilon_{zx}$
\label{t:irrep_strain}
\end{tabular}
\end{table}

\begin{table}
\caption{Fourth-order polynomial invariants of the phonon-strain
coupling.}
\begin{tabular}{cc}
$Q^1_4({\bf u},\epsilon)$ &
$(\epsilon_{xx} + \epsilon_{yy} + \epsilon_{zz})
( u^2_1 + u^2_2 + u^2_3 + u^2_4 + u^2_5 + u^2_6 )$ \\
\noalign{\vskip 3pt}
$Q^2_4({\bf u},\epsilon)$ &
$\bigl(2(u^2_1+u^2_2) - u^2_3-u^2_4-u^2_5-u^2_6\bigr) \times $\\ &
$(2\epsilon_{zz} - \epsilon_{xx} - \epsilon_{yy})+
3(u^2_3 + u^2_4 - u^2_5 - u^2_6)(\epsilon_{xx} - \epsilon_{yy})$ \\
\noalign{\vskip 3pt}
$Q^3_4({\bf u},\epsilon)$ &
$(-2 u_1u_2 - u_3u_4 - u_5u_6)(\epsilon_{xx} -\epsilon_{yy}) +$ \\ &
$(-u_3u_4 +u_5u_6)(2\epsilon_{zz} - \epsilon_{xx} - \epsilon_{yy})$ \\
\noalign{\vskip 3pt}
$Q^4_4({\bf u},\epsilon)$ &
$(u^2_1 - u^2_2)\epsilon_{xy} +(u^2_3 - u^2_4)\epsilon_{yz} +
      (u^2_5 - u^2_6)\epsilon_{xz}$
\label{t:inv_strain}
\end{tabular}
\end{table}

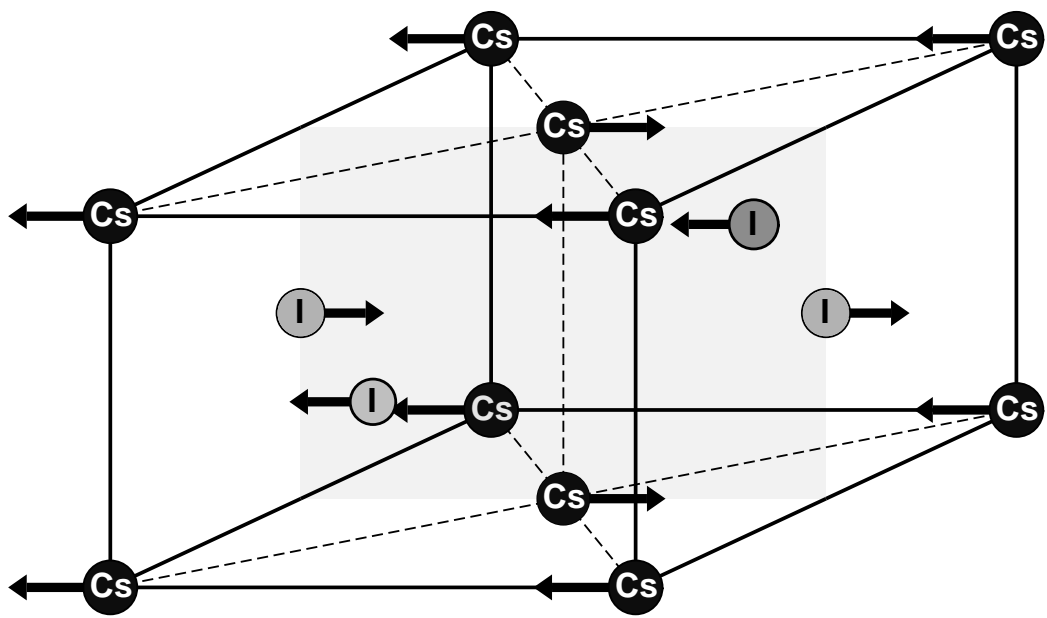
\begin{figure}
\caption{
Sketch of the CsI cell with the atomic displacements corresponding to an
$M_5^-$ acoustic phonon. The magnitude of Cesium displacements with respect to
Iodine displacements is exaggerated for clarity. The solid lines indicate the
cell of the distorted (orthorhombic) structure, while the cell of the
undistorted (cubic) structure is indicated by a dashed line. The shaded
area correspond to the (110) plane.}
\label{f:fig1}
\end{figure}

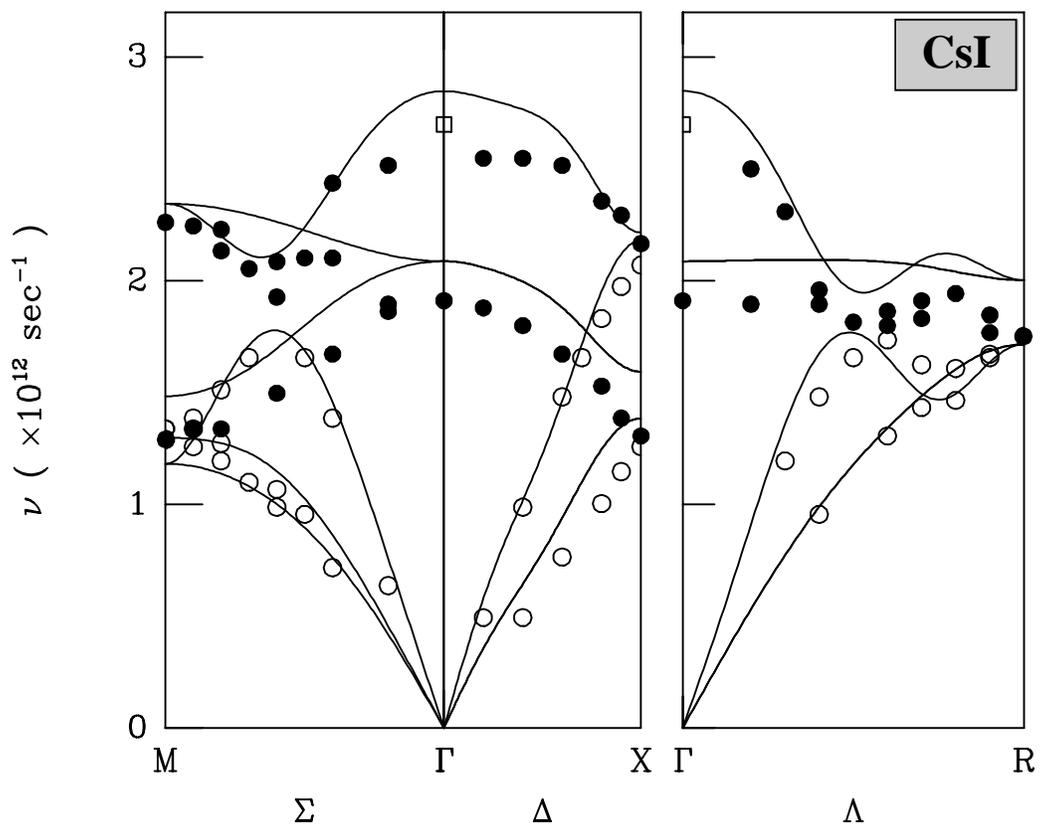
\begin{figure} \caption{Phonon dispersion relations along some
symmetry directions of the Brillouin zone in CsI at equilibriumi volume.
Experimental data indicated by open and black circles are from Ref.
\protect\onlinecite{Buhrer}; an infrared measurement for
$\protect\omega_{LO}(\protect\Gamma)$ is indicated by an open
square.\protect\cite{Lowndes} } \label{f:fig2} \end{figure}

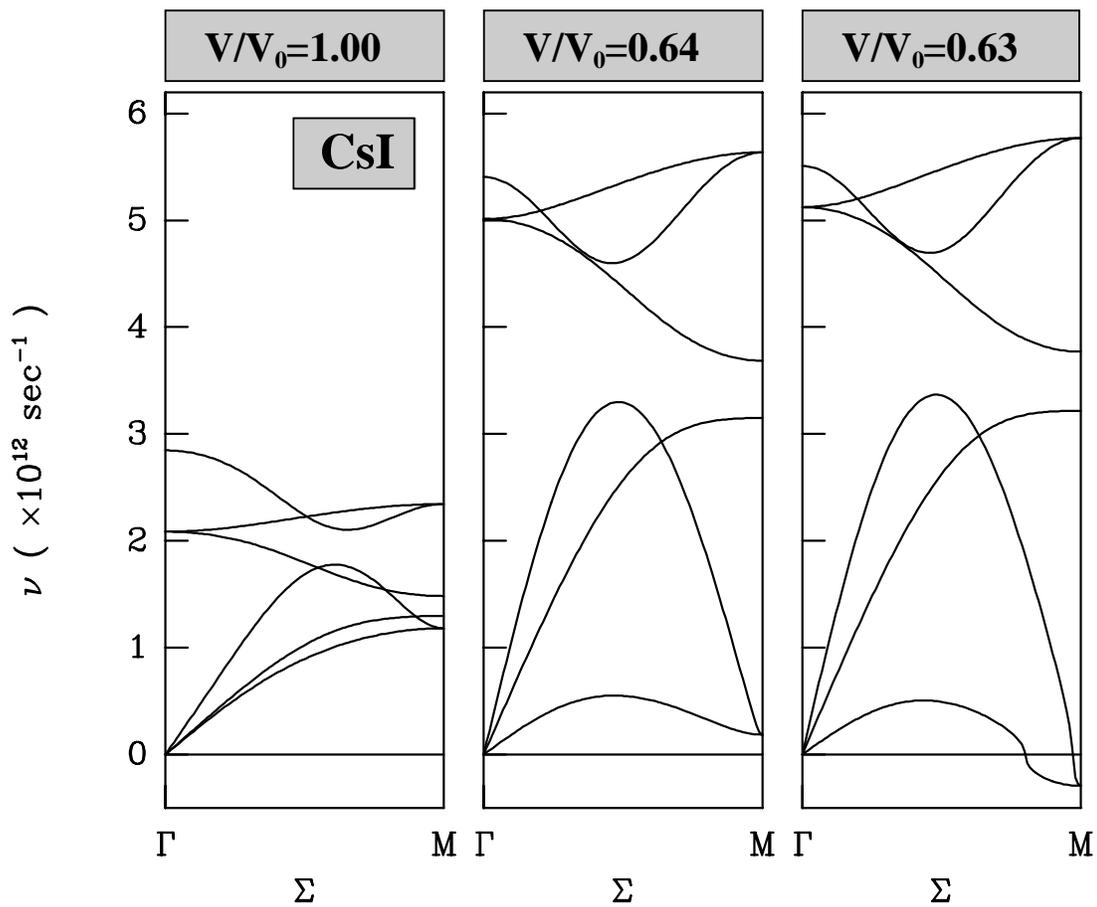
\begin{figure} \caption{ Phonon dispersion relations along the
$\protect\Sigma$ (110) for CsI at equilibrium volume and just above and
below the softening pressure of the $M_5^-$ acoustic mode. `Negative'
frequencies actually mean `imaginary' (i.e. negative squared
frequencies).} \label{f:fig3} \end{figure}

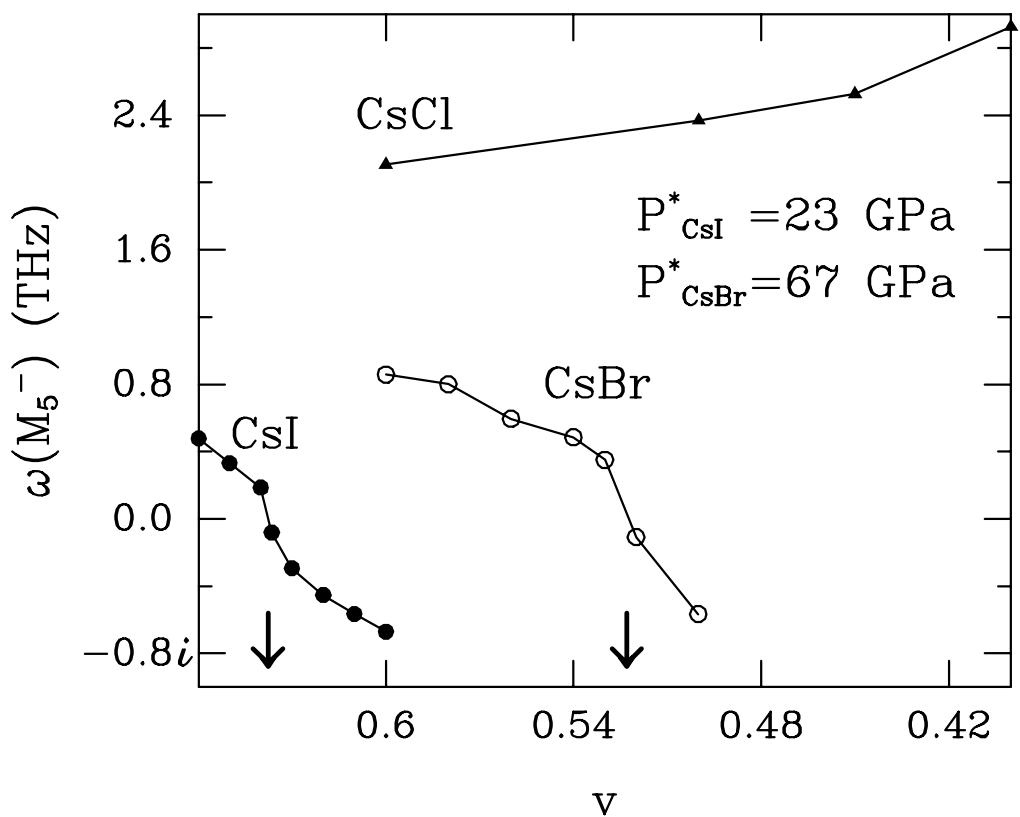
\begin{figure} \caption{Frequencies of the $M_5^-$ acoustic phonon as
functions of the molar volume for CsI, CsBr and CsCl. Arrows indicate
the softening volume. $P^*$ is the softening pressure.} \label{f:fig4}
\end{figure}

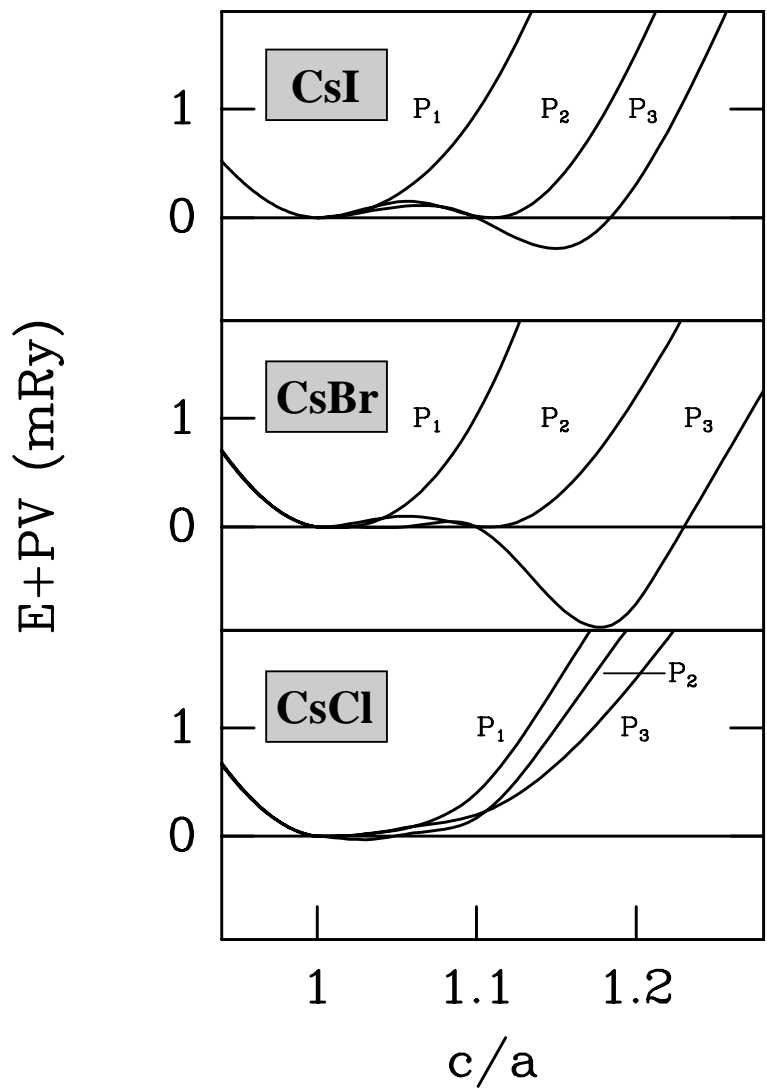
\begin{figure} \caption{Enthalpies {\it vs} $c/a$ for the three
compounds corresponding to a pressure $P_1 < P_{tet}; ~P_2\protect\approx
P_{tet}; ~ P_3 > P_{tet}$.} \label{f:tet1} \end{figure}

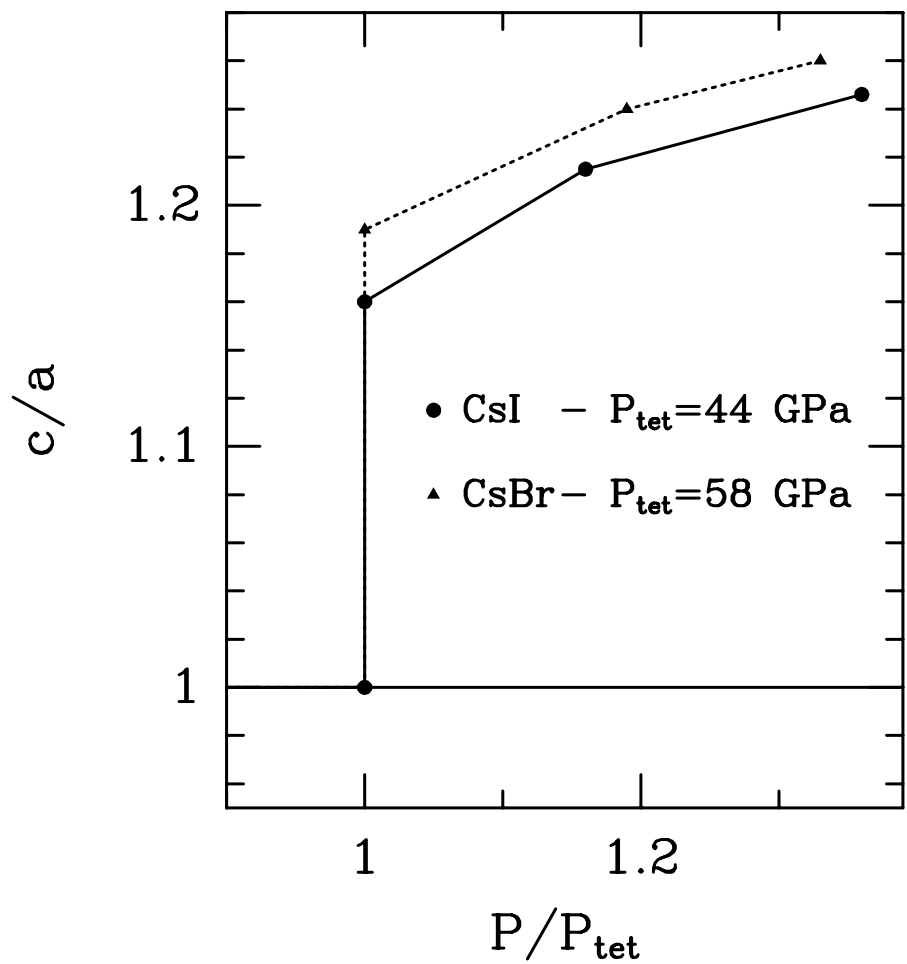
\begin{figure} \caption{$c/a$ parameter as a function of pressure for
CsI and CsBr.  $P_{tet}$ indicates the transition pressure to the tetragonal
phase.} \label{f:tet2} \end{figure}

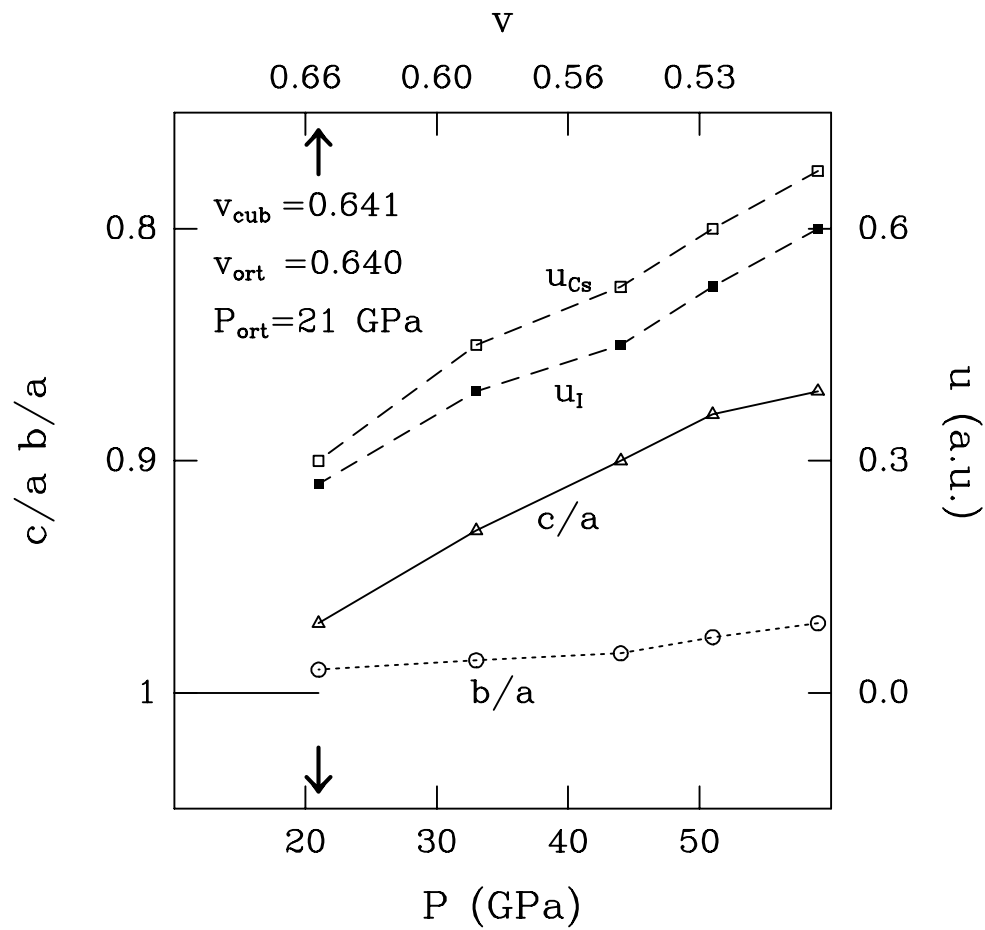
\begin{figure} \caption{Structural parameters of the ortho\-rhombic
phase of CsI as functions of different  values of the volume/pressure.
Arrows indicate  the transition pressure/volume. The volume
discontinuity at the transition is not visible on the scale of the
picture. $\protect{P_{ort}}$  indicates the transition pressure, while
$\protect{v_{cub}}$ and $\protect{v_{ort}}$ are respectively the volumes
of the high- and  low-symmetry phases at the transition (from Ref.
\protect\onlinecite{PRL}).} \label{f:fig5} \end{figure}

\begin{figure} \caption{Enthalpies of the orthorhombic and tetragonal
phases, relative to the enthalpy of the cubic phase, as functions of the
applied pressure in CsI and CsBr. $P_{ ort}$ and $  P_{tet}$ are the
transition pressures from the cubic to the tetragonal and to the orthorhombic
phases respectively.} \label{f:fig6} \end{figure}

\end{document}